\pdfoutput=1
\documentclass{article}

\usepackage{arxiv}

\usepackage{natbib}
\usepackage[utf8]{inputenc} 
\usepackage[T1]{fontenc}    
\usepackage{moreverb}
\usepackage{hyperref}
\usepackage{bibunits}

\usepackage{url, float}            
\usepackage{booktabs}       
\usepackage{amsmath}
\usepackage{pdfpages}

\usepackage{amsfonts}       
\usepackage{nicefrac}       
\usepackage{microtype}      
\usepackage{lipsum}		
\usepackage[nostruts]{titlesec}
\usepackage{graphicx}
\usepackage{doi}
\usepackage{enumitem}
\newlist{researchquestions}{enumerate}{1}
\setlist[researchquestions]{label*=\textbf{RQ\arabic*}}

\makeatletter

\newcommand*{\addFileDependency}[1]{
  \typeout{(#1)}
  \@addtofilelist{#1}
  \IfFileExists{#1}{}{\typeout{No file #1.}}
}
\@addtoreset{section}{part}

\makeatother

\newcommand\BibTeX{{\rmfamily B\kern-.05em \textsc{i\kern-.025em b}\kern-.08em
T\kern-.1667em\lower.7ex\hbox{E}\kern-.125emX}}

\titleformat{\part}[display]
{\normalfont\normalsize\bfseries\centering}{}{0pt}{}
\titleformat{\subparagraph}[runin]
{\normalfont}
{\thesection}{.5em}{}
\titlespacing{\section}{10pt}{10pt}{4pt}
\titlespacing{\subsection}{10pt}{10pt}{4pt}

\title{Mobility and Transit Segregation in Urban Spaces}

\author{ \href{https://orcid.org/0000-0002-5683-3023}{\includegraphics[scale=0.06]{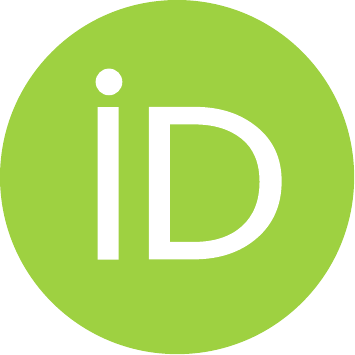}\hspace{1mm}Nandini Iyer}\\
	BioComplex Laboratory\\
	Computer Science Department\\
        University of Exeter\\
	Exeter, UK \\
	\texttt{niyer@biocomplexlab.org} \\
	\And
	\href{https://orcid.org/0000-0002-6479-6429}{\includegraphics[scale=0.06]{orcid.pdf}\hspace{1mm}Ronaldo Menezes}\\
	BioComplex Laboratory\\
	Computer Science Department\\
        University of Exeter\\
	Exeter, UK \\
	\texttt{r.menezes@exeter.ac.uk} \\
        \And
	\href{https://orcid.org/0000-0002-3927-969X}{\includegraphics[scale=0.06]{orcid.pdf}\hspace{1mm}Hugo Barbosa}\\
	BioComplex Laboratory\\
	Computer Science Department\\
        University of Exeter\\
	Exeter, UK \\
	\texttt{h.barbosa@exeter.ac.uk} \\
}
\date{}

\renewcommand{\shorttitle}{Mobility and Transit Segregation in Urban Spaces}

\defaultbibliography{ms}
\defaultbibliographystyle{unsrtnat}

\begin{document}
\maketitle

\begin{bibunit}
\part{}

\begin{abstract}
Segregation is a highly nuanced concept that researchers have worked to define and measure over the past several decades.  Conventional approaches tend to estimate segregation based on residential patterns in a static manner. In this work, we analyse socioeconomic inequalities, assessing segregation in various dimensions of the urban experience. Moreover, we consider the pivotal role that transport plays in democratising access to opportunities. Using transport networks, amenity visitations, and census data, we develop a framework to approximate segregation, within the United States, for various dimensions of urban life. We find that neighbourhoods that are segregated in the residential domain, tend to exhibit similar levels of segregation in amenity visitation patterns and transit usage, albeit to a lesser extent. We identify inequalities embedded into transit service, which impose constraints on residents from segregated areas, limiting the neighbourhoods that they can access within an hour to areas that are similarly disadvantaged. By exploring socioeconomic segregation from a transit perspective, we underscore the importance of conceptualising segregation as a dynamic measure, while also highlighting how transport systems can contribute to a cycle of disadvantage.
\end{abstract}


\section{Introduction}
Public transportation is a crucial component of urban environments, providing access to employment opportunities and amenities within a region. However, characteristics of transit systems, such as its urban layout and service frequency, can create pockets of transport deprivation, isolating particular neighbourhoods from conveniently accessing transit service \cite{nicoletti2022disadvantaged}. This is  dire for demographics that rely more on public transportation as their mode of transport \cite{hu2021left}. Lack of access to transport can impact how individuals perceive their activity space, by restricting or providing access to particular destinations \cite{tiznado2022freedom}. In this work, we analyse how US public transportation systems interact with segregation in residential and mobility landscapes, ultimately exploring whether they worsen or reduce existing levels of socioeconomic inequality. 

Understanding the effects of social integration has been at the forefront of sociological research for centuries, highlighting the many benefits of well-integrated communities, such as lower crime rates and better health outcomes \cite{kang2016inequality,scott2022place,asabor2022racial}. Much of the research on integration, and its counterpart, segregation, tends to focus on residential characteristics, assuming the individuals that one interacts with are likely to be from one's neighbourhood. Segregation is usually derived by comparing the empirical distribution of demographics in a region to an equally distributed version of the population. This is often followed by disclaimers that an equal distribution may not be ideal or just from a social perspective \cite{bruce2021urban}, offering a subtle nod to sociological studies that elucidate how homophily is not strictly a positive or negative force. Therefore, it is important to distinguish whether homophilic processes are occurring as a result of individual preferences or systemic constraints. Often, cities have highly concentrated pockets of immigrants, which, depending on the environmental context, has the potential to provide a sense of community to residents while also exposing the broader population to different cultures and practices \cite{mcpherson2001birds,peach1996good,miranda2020shape,klaesson2021ethnic}. Disparities arise when urban transport hinders individuals within these communities from travelling to other neighbourhoods in the region, allowing urban planning to shape segregation in mobility behaviour. 

The increasing availability of mobility data, such as Call Detail Records and credit card transactions, has enabled new forms of segregation to be conceptualised, addressing how segregation may persist in various urban domains \cite{phillips2021social,cagney2020urban}. These approaches extend conventional measures of segregation, assessing how mobility patterns can heighten or overcome segregation in activity spaces. Our work aims to expand on these studies, exploring whether segregation persists in various dimensions of the urban experience. We consider urban experience in terms of where individuals live, the amenities they visit, and how public transit systems facilitate their trips between the two locations. Thus, we pose the following research questions:

\begin{researchquestions}
  \item \textbf{What is the state of residential segregation in US cities?} To what extent are socioeconomic and racial segregation levels entangled? 
  \item \textbf{Does residential segregation persist in travel behaviour?} Does socioeconomic segregation in the mobility dimension impact socioeconomic groups differently?
  \item \textbf{How do transit systems contribute to existing levels of residential segregation?} Does transit service provide opportunities to travel to neighbourhoods of different economic backgrounds? Do inequalities in mobility patterns create disparities in the transit routes that economic groups would use?

\end{researchquestions}

We address the first research question by measuring segregation at a residential level. We estimate segregation using the Index of Concentration at the Extremes, 
evaluating the relationship between neighbourhoods' economic and racial segregation levels for 16 US cities \cite{booth2001prodigal}. 
We approach the second research question by defining segregation at the amenity level, leveraging SafeGraph mobility data and Census income distributions to define the composition of individuals visiting an amenity. Thus, we can analyse how neighbourhoods levels of segregation change based on its residential composition and the amenities to which its residents are travelling. Finally, we construct transit-pedestrian networks to evaluate how public transportation can mitigate or worsen existing levels of segregation. For different travel time thresholds, we assess the socioeconomic profile of neighbourhoods that can be accessed through public transit compared to driving. Moreover, we implement a stochastic model to estimate the level of segregation experienced while using the transit network, to ultimately investigate whether disparities in mobility destinations can contribute to particularly segregated segments in the transit network. Our results show that segregated low-income neighbourhoods tend to exhibit less segregated mobility behaviour than segregated high-income neighbourhoods. Ultimately, our findings highlight disparities in how transit systems provide access to different neighbourhoods, while revealing that levels of residential segregation linger in other aspects of urban life.

\section{Background}
Residential segregation has been in the spotlight of sociological and urban studies for centuries, with research consistently churning out new methods for conceptualising and measuring how sociodemographic groups share spaces \cite{massey1988dimensions}. In this work, we quantify segregation using the Index of Concentration at the Extremes (ICE), which reconciles the diverging studies of concentrated affluence and concentrated poverty, to ultimately interpret them as one continuum \cite{booth2001prodigal}. This is achieved by comparing how many households or individuals from the most deprived and privileged groups share the same residential area. For a given region $i$, with $T_i$ total households, $A_i$ affluent, or privileged households, and $P_i$ households in poverty, the ICE can be calculated as follows: 

\begin{equation}
ICE_{i} = \frac{A_i - P_i}{T_i},
\label{eq:ICE}
\end{equation}

Values can range from -1 to 1, reflecting extreme concentration of disadvantaged and privileged households, respectively. Thus, the ICE can capture levels of imbalance given the sociodemographic composition of a region. Section \ref{si:ice_seg_corr} in the Supplementary Materials illustrates how ICE values correspond to measures of Dissimilarity, Exposure, Mutual Information, and social distance. We move forward, using ICE, as it clearly distinguishes between segregation of the most and least privileged demographics. 

\subsection{Human Mobility}

Mobility offers opportunities for overcoming the segregation that residential mechanisms, such as the housing market, impose on individuals. The prevalence of GPS data and Call Detail Records has allowed for high-resolution analysis of travel behaviours \cite{barbosa2018human}. Studying mobility trajectories has highlighted the predictability of travel patterns, and, in turn, resulted in mobility models that can reproduce such behaviour \cite{zipf1946p,simini2012universal,song2010modelling,pappalardo2015returners}. Moreover, access to descriptive mobility data provides insight in whether one's mobility patterns are influenced by their economic standing \cite{barbosa2021uncovering}. 
Moro et al. use mobility data to build an extension of the Exploration and Preferential Return model, which identifies an association between experienced income segregation and individuals' level of place exploration \cite{moro2021mobility}. In doing so, they demonstrate how experienced segregation is related to residential characteristics and amenity visitation patterns. Various research in mobility inequalities has identified exacerbated levels of income segregation following natural disasters and relationships between income inequality and segregation in various countries, hindering social mobility \cite{yabe2020effects,nieuwenhuis2020does}. Moreover, studies have found persisting segregation in activity spaces for disadvantaged neighbourhoods \cite{abbasi2021measuring,wang2018urban}. These studies highlight the benefits of considering segregation from both a residential and a mobility-based perspective. 

Transport poverty, a field that has a large overlap with human mobility, has been gaining traction in the urban planning realm. It is often conflated with terms such as transport affordability, mobility poverty, accessibility poverty, but Lucas argues that these terms are merely a subset of what transport poverty represents \cite{lucas2016transport}. Inequalities in transport systems, and the types of amenities and neighbourhoods they provide access to, is important to consider as it can impact the level of choice that disadvantaged groups have when using transit \cite{tiznado2022freedom}. Moreover, an analysis of Colombian cities found more affluent areas to benefit from better transit coverage and employment opportunities, identifying longer average travel times for individuals in poorer neighbourhoods \cite{arellana2021urban}. We extend these analyses, by exploring how US transit systems relate to residential and amenity segregation.

\section{Methods and Data}
This work combines census, mobility and transit data to analyse how transportation systems intersect with segregation in different aspects of urban life. We first define the state of residential segregation, using US Census data. Then, with anonymised mobility patterns from SafeGraph, we define segregation levels for amenities, based on the socioeconomic composition of its visitors. Finally, drawing upon open source resources, such as The Mobility Database, OpenStreetMap, and  UrbanAccess, we construct transit networks to identify disadvantages within the system, analysing both the transport service and experience of using transit routes as potential sources of inequality.

\subsection{Sociodemographic Data for Measuring Segregation}

Typical measures of segregation focus on inequalities experienced in residential areas. In this work, using the ICE metric from Equation \ref{eq:ICE}, we define segregation with respect to the socioeconomic concentration in three urban contexts: (a) residential (b) amenities and (c) public transit. We define residential segregation by drawing upon the 2020 American Community Survey 5-Year Estimates (ACS), provided by the US Census Bureau. Household income distributions, from Table B19001 of the ACS, inform the economic composition of a neighbourhood, at the Census Block Group (CBG) level \cite{us2020acs}. CBGs are the smallest spatial unit that the Census Bureau publishes data for, typically consisting of 600 to 3,000 individuals. Thus, segregation levels of CBGs are calculated using these income distributions. To compare levels of economic and, we use Table B02001 from the ACS, which captures the racial distribution of individuals in a CBG. Considering US history, and its persistent discrimination against the Black population, we select Black and White racial groups to represent the extremes in the context of racial segregation. \cite{franklin1956history}.

\subsection{Mobility Data}\label{sssec:amenity_data}
In order to capture more dynamic forms of segregation, we draw upon mobility data sources to better understand the role of mobility in overcoming residential segregation. Our mobility data is sourced from SafeGraph
, a data company that aggregates anonymised location data from numerous applications in order to provide insights about physical places, via the SafeGraph Community \cite{safegraph}. To enhance privacy, SafeGraph excludes census block group information if fewer than two devices visited an establishment in a month from a given census block group. The SafeGraph Weekly Patterns data provide visitation counts, on a weekly level, to amenities across the US, along with the distribution of CBGs from which the visitors came. Home CBGs are defined by SafeGraph, using users’ locations from 18:00 to 07:00 over a six-week time frame. For this analysis we use amenity visitations from January 2021 to characterise mobility flows from CBGs to amenities. With this data we can estimate the volume of trips between any pair of CBGs in a city. Moreover, we can define a CBG by the types of destinations to which its residents travel.

\subsection{Transportation} \label{sssec:transport_data}
We leverage General Transit Feed Specification (GTFS) data, from the Mobility Database, to build public transportation networks for 16 US cities \cite{mobilitydatabase}. GTFS refers to a data format for publishing public transit schedules and routes, for various forms of transit. Using UrbanAccess, an open source tool that combines transit networks, defined by GTFS data, with pedestrian networks, built using OpenStreetMap, to create a bipartite, transit-pedestrian network \cite{blanchard2017urbanaccess}. Nodes are either transit nodes, representing public transit stops, or pedestrian nodes, reflecting street nodes from the OSM network. The edges are weighted by travel time in minutes. Further details regarding the construction of the transit-pedestrian network can be found in \ref{si:transit_nx} of the Supplementary Materials.

Moreover, we use Open Source Routing Machine (OSRM) to generate driving times and distances between any pair of coordinates in a city, given an OpenStreetMap (OSM) extract. OSRM is a high-performing routing engine that integrate well with OSM to find shortest paths on a road network. Driving times serve as a baseline for travel time, allowing us to compare how much longer trips take using transit than by driving. While cars and public transit vehicles both use road networks, transit vehicles must adhere to determined schedules and routes, while cars have much fewer constraints as to how they can traverse the road network. Thus, driving terms are useful for understanding the impedance, in terms of travel time, of using the transit system.

\section{Residential Segregation} \label{sec:res_seg}
We begin by exploring the state of residential segregation in 16 US cities. We take a closer look at the relationship between racial and socioeconomic composition to develop a better understanding of the residential landscape throughout the US. In doing so, residential segregation acts as a baseline, to which we can compare segregation levels in the other urban dimensions we consider in the coming sections.

\subsection{Sociodemographic Residential Segregation\label{ssec:raceVSincome_res_seg}}
\begin{figure}
    \centering
    \includegraphics[width=0.6\textwidth]{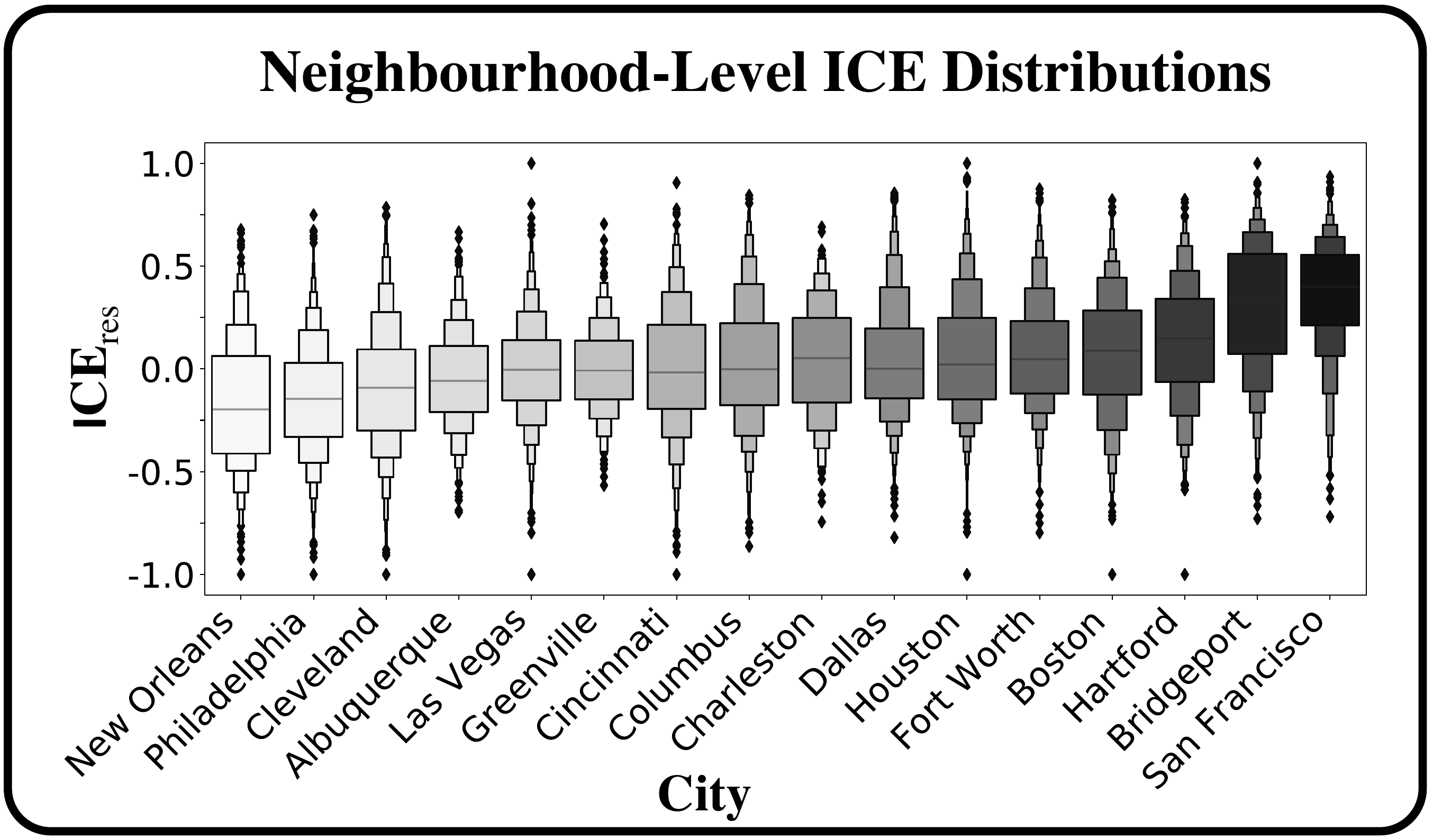}
    \caption{Socioeconomic residential segregation in 16 US cities, calculated using ICE. Each box plot reflects the distribution of ICE values in census block groups, for a given city.}
    \label{fig:ICE_distr_allCities}
\end{figure}

The ACS data provides the number of households belonging to each of the 16 income brackets, for every census block group. We denote the lower three income brackets, that earn less than \$20,000 per year, as households in poverty. Meanwhile, the upper three income brackets indicate affluent households, which earn more than \$125,000 per year. We define these brackets as high income households. Middle class households are reflected by the middle 10 income brackets. Thus, we can categorise each of the 16 income brackets into 3 income classes: low-income, middle-income, and high-income. Using these cutoffs to define income classes is common practice when measuring ICE at the CBG-level \cite{larrabee2022racialized,bishop2021structural,wallace2019privilege}. However, Section \ref{si:ice_seg_cutoffs} in the Supplementary Materials conveys how ICE distributions would change within each city if we were to shift these income-bracket cutoffs. With this in mind, we define $ID_{n,i}$ as the the number of households belonging to an income class, $i$ in CBG $n$. We combine this data, with our measure of segregation (Eq. \ref{eq:ICE}) to define residential segregation for a census block group, $n$:

\begin{equation}
    ICE_{res}(n) = \frac{ID_{n,hi}-ID_{n,lo}}{\sum_{i \in I} ID_{n,i}}
    \label{eq:ICE_res}
\end{equation} 

where $I$ reflects each of the three income classes (lo, mid, and hi). Conceptually, we are computing the extent to which households from affluent and poor households live in the same neighbourhood. 
We emphasise that the overall socioeconomic composition varies across cities, with San Francisco housing a far larger number of affluent households, as indicated through Figure \ref{fig:ICE_distr_allCities}. Each box plot in Figure \ref{fig:ICE_distr_allCities} reflects the distributions of socioeconomic ICE values for all neighbourhoods in the corresponding city. Accordingly, when evaluating segregation within the context of one city, it is important to consider ICE values with respect to the overall composition of the city. 

\begin{figure}
    \centering
    \includegraphics[width=\textwidth]{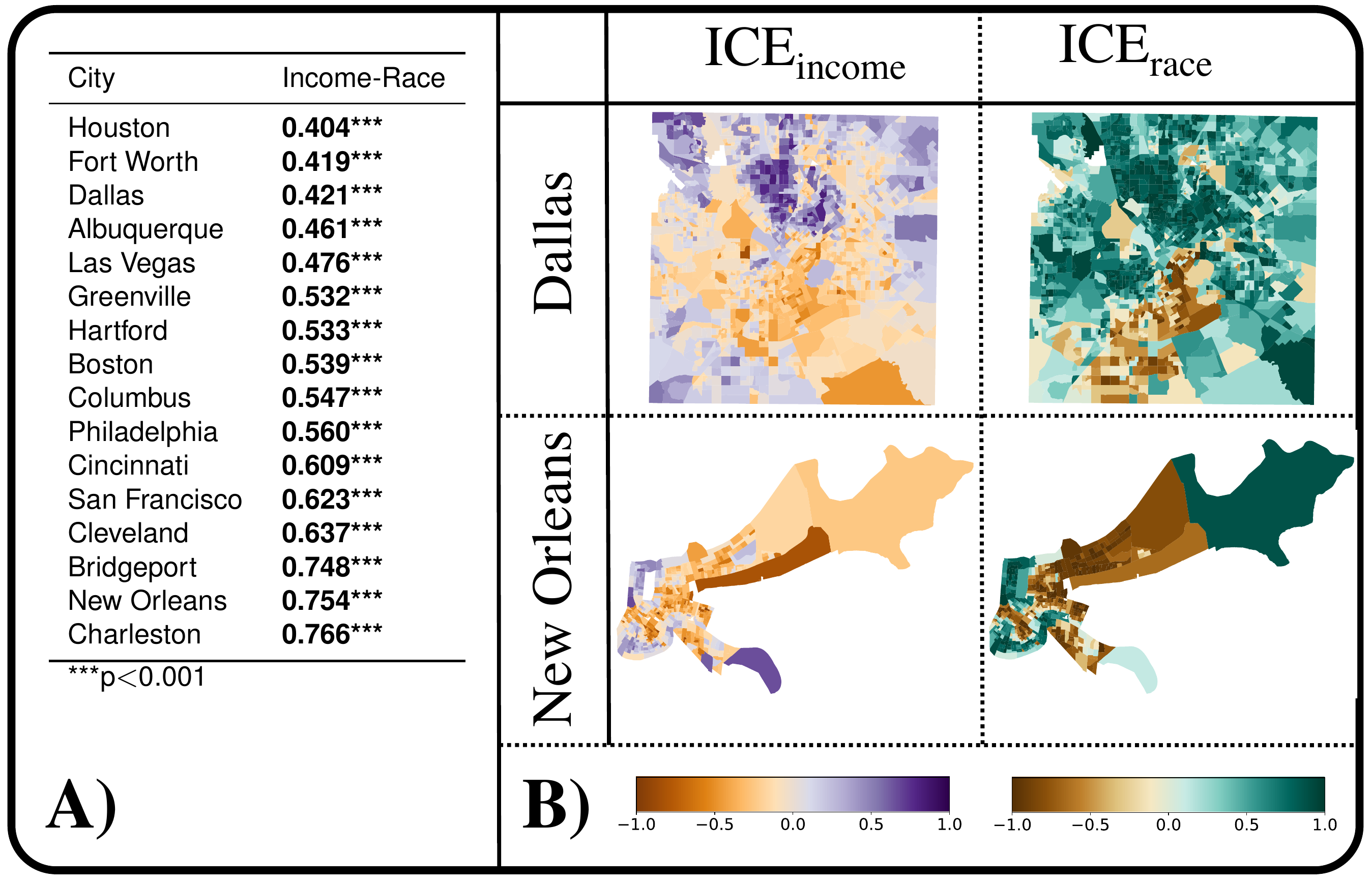}
    \caption{A) Pearson correlation coefficients, emphasising the relationship between economic and racial segregation in 16 US cities. B) Maps illustrating the entangled nature of economic and racial segregation for Dallas and New Orleans. Orange and purple reflect low and high-income concentrations, respectively. Brown and blue capture a high residential concentration of Black and White residents, respectively.}
    \label{fig:race_income_ICE_corr}
\end{figure}

Similarly, we compute the residential segregation of Black and White residents using Equation \ref{eq:ICE}, such that $A_i$ and $P_i$ represent the number of White and Black residents, respectively, in a neighbourhood $i$. Accordingly, values of -1 indicate a high concentration of Black residents while +1 ICE levels reflect a large share of White residents. The second column in Panel A of Figure \ref{fig:race_income_ICE_corr} captures the relationship, using Pearson correlation coefficients, between a neighbourhood's economic and racial segregation level, for 16 US cities. A positive correlation indicates that neighbourhoods, or CBGs, with a large share of affluent households also have a high concentration of white residents. While the degree of correlation between the two demographic types varies across cities, we observe that all cities do have a positive, significant correlation between racial and economic segregation, indicated by the asterisks. 

Panel B in Figure \ref{fig:race_income_ICE_corr} visualises the spatial landscape of economic and racial segregation in cities with lower and higher correlation coefficients (Dallas and New Orleans, respectively). Comparing $ICE_{income}$ to $ICE_{race}$ in Dallas reveal that areas with a large concentration of low-income residents are not directly translatable to highly segregated Black or highly segregated White neighbourhoods. Panel B indicates that concentrated poverty exists for both of the considered racial groups, however Dallas' positive correlation coefficient implies that neighbourhoods with  lower still have more concentratio. On the other hand, New Orleans shows strong associations between neighbourhoods that are low-income and segregated (orange CBGs) and those that are largely composed of Black residents (brown CBGs). By comparing CBG-level segregation measures, with respect to socioeconomic and Black-White composition, we illustrate how residential patterns for income and racial groups can be intertwined. Measuring segregation levels characterises neighbourhoods based on the sociodemographic composition of its residents. However, accessibility can provide a glimpse of the diversity of amenities to which a neighbourhood's residents can travel. Moving forward, we focus on analysing income segregation, however, the proposed methodology can be applied to explore whether racial segregation persists in urban dimensions.

 \section{Mobility Segregation}\label{sec:mob_seg}

In this section, we extend our analyses of residential segregation in US cities, to explore whether residents from segregated neighbourhoods tend to exhibit segregation in their amenity visitation patterns. To do so, we define the segregation of amenities, based on the socioeconomic composition of its visitors during January 2021. Then, we explore whether residents from the most segregated neighbourhoods mitigate their residential segregation by traveling to amenities that have visitors from different economic backgrounds. 
This approach allows us to understand mobility differences between highly-segregated high-income neighbourhoods and highly-segregated low-income neighbourhoods. 

\subsection{Segregation at the Amenity Level}\label{ssec:amenity_seg}
We begin analysing segregation from a human mobility perspective, by measuring segregation based on the socioeconomic makeup of an amenity's visitors. We define $ID'$ as the normalised form of the income distribution, $ID$, defined by the ACS data:
\begin{equation}
    ID'_{n,i} = \frac{ID_{n,i}}{\sum_{i' \in I} ID_{n,i'}} \textit{, where } I = \{lo, mid, hi\}
    \label{eq:norm_ID}
\end{equation}

$ID'_{n,i}$ defines the fraction of households in CBG $n$ that belong to income class $i$. The SafeGraph data provide information regarding the mobility flow from CBGs to amenities, while the ACS data denote the socioeconomic composition of a CBG. Using the SafeGraph amenity visitations from the Weekly Pattern data set, we construct an $|\textbf{N}| \times |\textbf{A}|$ mobility matrix, $\textbf{M}$, for January 2021, where $\textbf{N}$ and $\textbf{A}$ are the set of CBGs and amenities in a city, respectively. $\textbf{M}_{n,a}$ reflects the number of trips from CBG $n$ to amenity $a$ during the month. Combining the two sources, we can estimate the socioeconomic composition of mobility flows between CBG-amenity pairs, by performing a weighted sampling of $ID'$, indicated by the $\sim$ symbol:

\begin{equation}
    \textbf{C}_{n,a} = \{C_{n,a,1}...C_{n,a,v}\},
    \quad \textit{where }C_j  \sim ID'_n,\ 
    v = \textbf{M}_{n,a}
    \label{eq:mob_comp}
\end{equation}

Here, $C_j$ represents an individual from CBG $n$, who visits amenity $a$, and belongs to an income class $i \in I$, which is sampled from $ID'_n$. Due to the level of anonymisation in the SafeGraph data, this method of sampling assumes that individuals in a neighbourhood have an equal likelihood of traveling to each amenity. The segregation level of individuals visiting an amenity, then, is determined by a neighbourhood's socioeconomic distribution and the volume of its residents that travel to the amenity. Even under this assumption, we still identify segregation in visitation patterns. Moreover, to account for the stochastic nature of this approach, we perform the weighted sampling 100 times, where $C_{n,a}^x$ reflects the economic composition of  visitors from CBG $n$ visiting amenity $a$, during the $x^{th}$ iteration. 

We can modify $\textbf{C}_{n,a}^x$, to solely reflect the socioeconomic composition of each amenity such that $\textbf{C}_a^x = \{\textbf{C}_{n,a}^x \ | \ n \in N\}$ and $|\textbf{C}_a^x| = \textbf{M}_a$. Each element in $\textbf{C}_a^x$ resembles the income group of one of amenity $a$'s visitors for the $x^{th}$ iteration. Thus, $\textbf{C}_{a,i}^x$ captures, for a given iteration, the set of individuals visiting amenity $a$, that are in an income group $i$, where $i$ reflects a low, middle, or high-income group, as defined in Residential Segregation Section. We can define the level of segregation, in terms of mobility patterns, at amenity $a$ with the following equation:

\begin{equation}
    ICE_{amenity}(a) = \frac{1}{100} \sum\limits_{x=1}^{100}\frac{|\textbf{C}_{a,hi}^x| - |\textbf{C}_{a,lo}^x|}{|\textbf{C}_{a}^x|} 
    \label{eq:amenity_ICE}
\end{equation}

Equation \ref{eq:amenity_ICE} computes the average segregation of an amenity's visitor composition using ICE, such that the economic makeup of visitors is determined through a stochastic, weighted sampling with respect to visitation frequency and the socioeconomic characteristics of visitors' origins.
 Having used the Index of Concentration at the Extremes and visitation patterns to measure amenity segregation, we define mobility segregation from the perspective of residents in a neighbourhood, based on the level of segregation they experience, on average, at the amenities they visit. We refer to this measure as \textbf{traveler amenity segregation} (TAS) as it captures segregation based on residents' mobility patterns. Traveler amenity segregation, for a neighbourhood, only considers segregation at the amenities its residents visit. Then, the TAS of a neighbourhood, $n$, is computed as the weighted average of these amenities' segregation levels, with respect to the frequency with which its residents visit the amenities:
\begin{equation}
    {TAS}(n) = \frac{\sum_{a \in A} \textbf{M}_{n,a}*ICE_{amenity}(a)}{\sum_{a \in A} \textbf{M}_{n,a}}
    \label{eq:TAS}
\end{equation}

where $\textbf{M}_{n,a}$ is the number of visits from CBG $n$ to amenity $a$, as outlined in Equation \ref{eq:mob_comp} and $A$ is the set of all amenities in a city. Thus, the amenities that are more frequently visited will play a bigger role in characterising the average segregation of amenities to which residents travel. 

\subsection{Mobility Segregation and Residential Segregation}\label{ssec:TAS_vs_Res}

\begin{figure}
    \centering
    \includegraphics[width=0.6\textwidth]{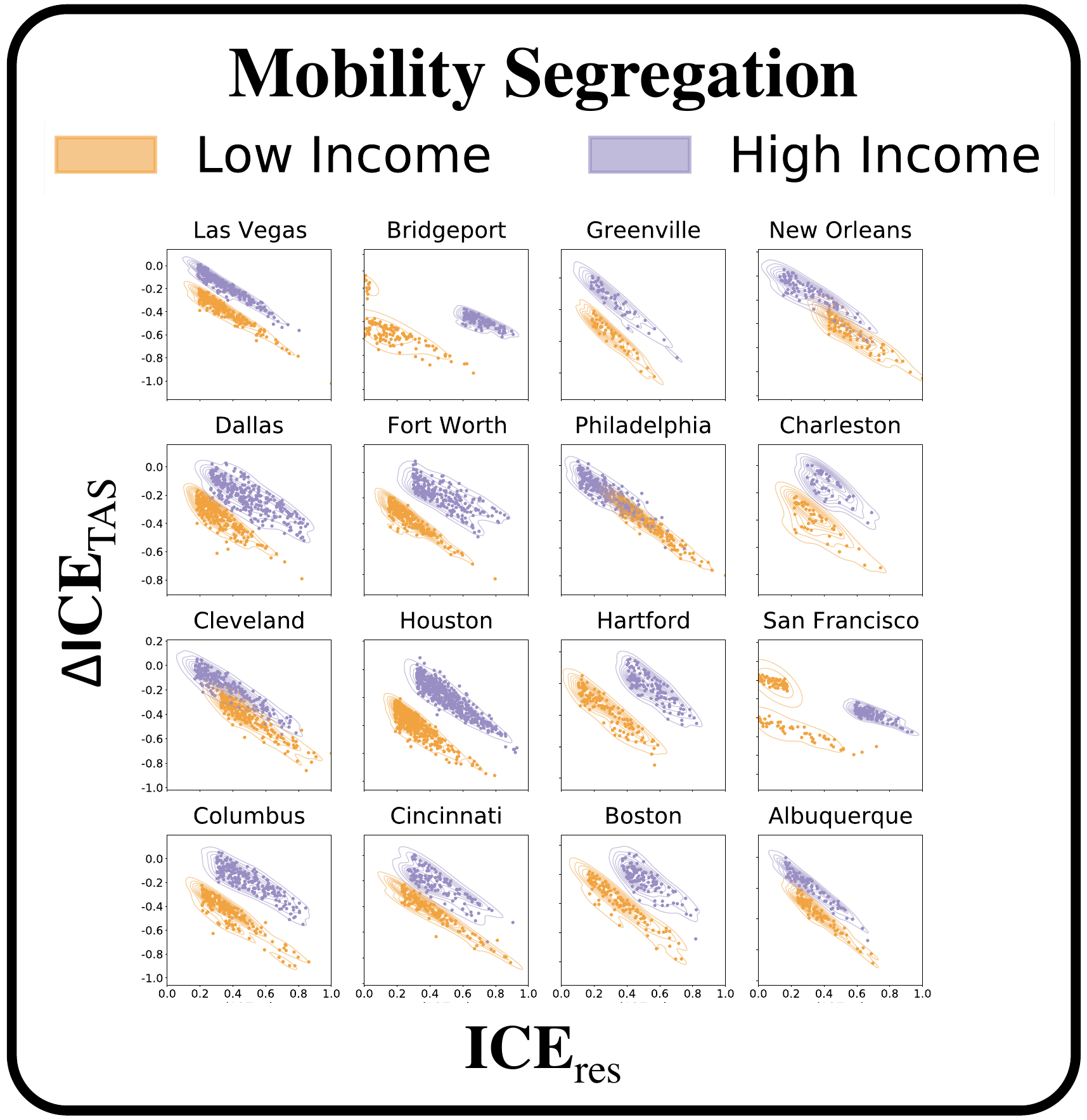}
    \caption{Differences in segregation levels in the residential and mobility domain, where each scatter plot resembles one of the 16 US cities. The x-axis depicts the magnitude of residential segregation, for the highly-segregated, high-income neighbourhoods (purple) and the highly-segregated, low-income CBGs (orange). The y-axis shows the direction and magnitude of change in a CBG's mobility segregation level, compared to its residential segregation level.}
    \label{fig:traveler_amenity_seg}
\end{figure}

From a broader perspective, TAS aims to depict how individuals experience segregation in the mobility dimension, by considering their visitation patterns and the segregation levels of the amenities they visit. 
With this in mind, we examine the relationship between residential segregation and traveler amenity segregation, illustrated in Figure \ref{fig:traveler_amenity_seg}. We can calculate how much a segregation value, for a given neighbourhood, changes between two urban dimensions, given it's original $ICE$ value, $x$, and its $ICE$ value in a different dimension, $y$:
\begin{equation}
    \Delta ICE(x, y) =
                \begin{cases}x - y, & \text{if } x < 0\\
                - (x - y), & \text{if } x \geq 0
                \end{cases}
    \label{eq:delta_ICE}
\end{equation}

$\Delta ICE$ ranges from -2 to 2, where negative values signal a decrease in segregation levels. If a neighbourhood has a large concentration of high income residents, such that its $ICE_{res}$ is positive, having a lower $ICE$ value in the mobility dimension is indicative of a decrease in segregation levels when shifting from the residential and mobility dimension. However, a decrease in amenity segregation levels for neighbourhoods with negative $ICE_{res}$ values, reveals that its residents are traveling to amenities that have a larger concentration of low-income individuals, than compared to their neighbourhood's residential composition. Such instances signify an increase in segregation from the residential to mobility dimension. Thus, the two cases account for the sign of the original $ICE$ value. We denote the difference between $ICE_{res}(n)$ and $TAS(n)$ in a neighbourhood $n$, as $\Delta ICE_{TAS, n}$, which expresses how segregation levels change between the residential and mobility dimension.

For each city, we split neighbourhoods, using residential ICE values, into five, equally sized segregation groups: (1) highly-segregated; low-income (\textbf{HS-Lo}) (2) moderately-segregated; low-income (\textbf{MS-Lo}) (3) less-segregated (\textbf{LS}) (4) mildly-segregated; high-income (\textbf{MS-Hi}) and (5) highly-segregated; high-income (\textbf{HS-Hi}). Since we create the segregation groups with respect to each city, neighbourhoods belonging to the highly-segregated, low-income group do not necessarily have high segregation magnitudes. Rather, they reflect neighbourhoods that have a high concentration of low-income households, relative to the segregation distribution of that city. Figure \ref{si:seg_groups} in the Supplementary Materials shows the distribution of $ICE_{res}$ values for each segregation group in a city. We focus on the highly segregated low-income and high-income neighbourhoods, represented in Figure \ref{fig:traveler_amenity_seg} by the orange and purple points, respectively. The x-axis captures the magnitude of residential segregation ($|ICE_{res, n}|$), which allows us to compare the most segregated neighbourhoods in a city within the same frame of reference. 

The consistently negative slopes in Figure \ref{fig:traveler_amenity_seg} can be attributed to the positive correlations between residential segregation and $TAS$, shown in Table \ref{si:res_mob_corr} of the Supplementary Materials. Figure \ref{fig:traveler_amenity_seg} reveals that the majority of CBGs tend to have negative $\Delta ICE_{TAS}$ values. These smaller values suggest a decreased level of segregation when comparing the residential segregation individuals experience in their neighbourhood, to their $TAS$ values that measure mobility segregation. 

The key takeaway of Figure \ref{fig:traveler_amenity_seg}, however, is that, for most cities, the highly-segregated, low-income neighbourhoods exhibit smaller values of $\Delta ICE_{TAS}$ than their high-income counterparts. This can be observed by considering how each group's points are distributed along the y-axis. Yet, two cities emerge as exceptions to this trend. A subset of highly-segregated, low-income neighbourhoods in San Francisco and Bridgeport exhibit distinctive patterns in which they experience an increase in $\Delta ICE_{TAS}$, pointing to an increase in mobility segregation levels, despite already having high levels of residential segregation. In this manner, Figure \ref{fig:traveler_amenity_seg} reveals that, generally, segregated low-income neighbourhoods tend to mitigate their residential segregation level by traveling to amenities with a much different economic composition, than compared to segregated high-income neighbourhoods. These findings emphasise the importance of considering segregation from various urban layers, as mobility can be used as a means to decrease the overall segregation that one experiences. While, we find the mobility segregation occurs in lower magnitudes than residential segregation, this section highlights how segregation continues to exist when considering amenity visitation patterns, finding strong associations between the two domains. 

\section{Transport Segregation}\label{sec:Transport_Seg}
Having demonstrated the role that segregation plays in the residential and mobility facets of the urban experience, we move on to consider the intersection between segregation and public transportation systems. In this section we leverage public transit networks to analyse how structural properties of transportation systems coincide with the residential landscape. Employing the SafeGraph mobility data, we model potential transit use to estimate the level of segregation one would experience while using transit to satisfy her mobility demands. Due to the computational complexity of stochastically modeling transit use, we use five of the 16 cities as an example for how levels of structural and experiential segregation can be assessed in the transit system. We specifically choose New Orleans, Philadelphia, Cincinnati, Dallas, and San Francisco as the 5 focal cities, as each city spans different parts of the socioeconomic residential composition, as depicted in Figure \ref{fig:ICE_distr_allCities}. 

\subsection{Structural Transport Segregation}\label{ssec:Structural_Transit_seg}

\begin{figure}
    \centering
    \includegraphics[width=0.8\textwidth]{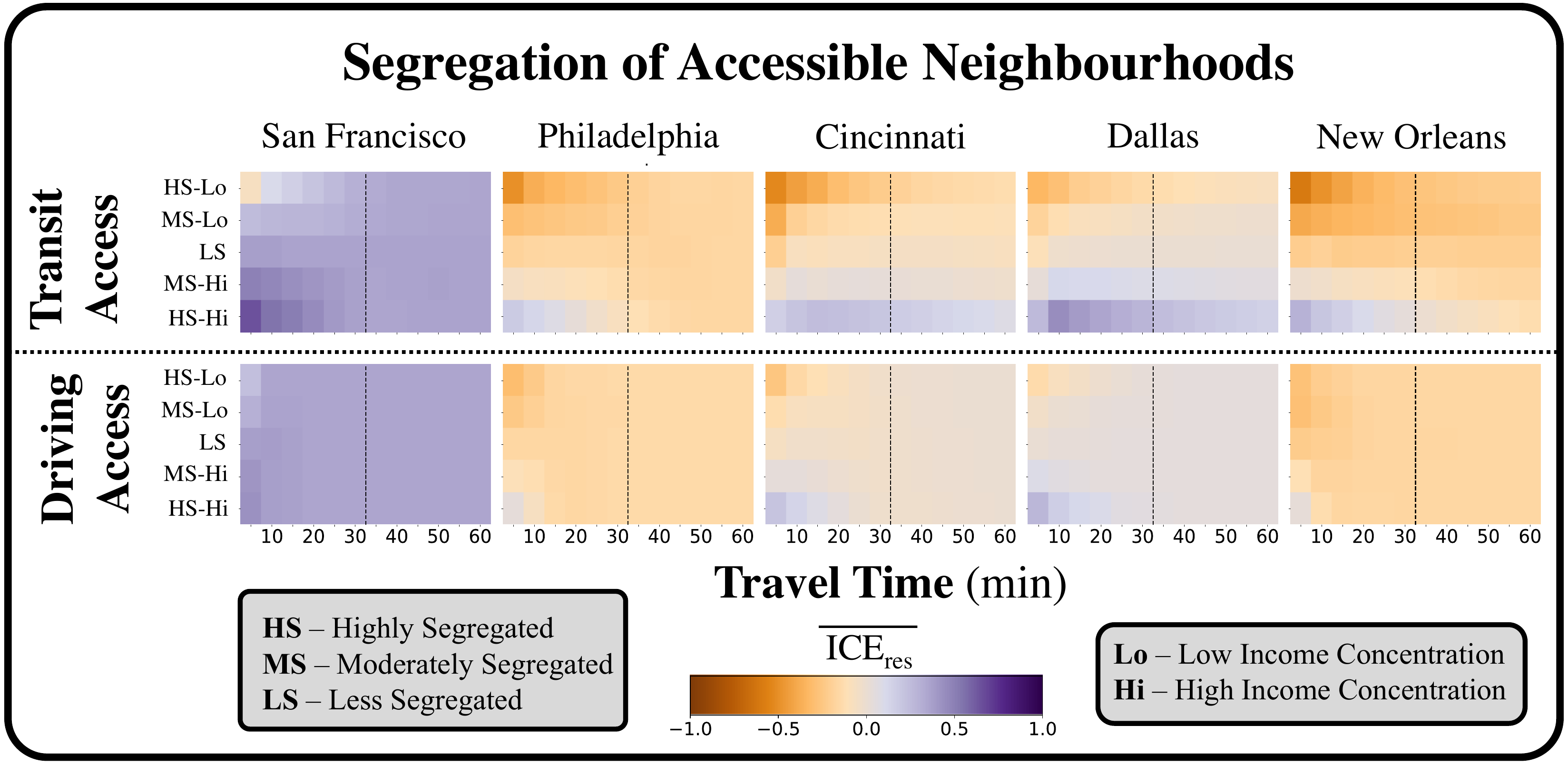}
    \caption{Average residential segregation level of neighbourhoods that are reachable within a given travel time, by public transit (top row) and car (bottom row), for 5 US cities. The y-axis shows neighbourhood accessibility for different segregation groups, while the x-axis defines different time thresholds. Orange cells reflect accessibility of neighbourhoods that are more segregated, with a concentration of low-income households. Purple cells capture transit service to neighbourhoods with a higher income concentration.}
    \label{fig:structural_transit}
\end{figure}

We assess segregation in the context of transport by, first, examining how transit systems serve neighbourhoods with various segregation levels. To do so, we consider travel times between every possible pair of neighbourhoods in a city. For every CBG pair, $(n_o, n_d)$ we sample 100 points from the origin census block group, $\{n_{o}^{1}...n_{o}^{100}\}$, and 100 from the destination CBG, $\{n_{d}^{1}...n_{d}^{100}\}$. Since edges in the transit-pedestrian networks are weighted by travel time, we can find the shortest path length on the transit-pedestrian networks between points $n_{o}^{i}$ and $n_{d}^{i}$, for $i \in \{1...100\}$. We calculate the corresponding driving times for the same set of coordinate pairs, using Open Source Routing Machine. As a result, we can measure the average time it takes to travel between two neighbourhoods in a city, using public transportation or a car. 
Accordingly, we denote matrices $T_{transit}(n_1, n_2)$  and $T_{driving}(n_1, n_2)$ to reflect the average travel time between neighbourhoods $n_1$ and $n_2$, when using public transport and cars, respectively. 
We use these matrices to evaluate which neighbourhoods can be accessed within a travel time threshold. Specifically, we use the previously defined segregation groups, which are derived by partitioning a city's neighbourhoods into 5 equally sized groups based on their residential segregation levels. As follows, $\textbf{N}_s$ refers to the set of neighbourhoods in segregation group $s$. Given a travel time threshold, $t$, and a CBG, $n_s\in\textbf{N}_s$, we use $T_{transit}$ to define $\textbf{N'}_{n_s}$ as the neighbourhoods that can be reached by $n_s$ within $t$ minutes. Subsequently, we calculate the average segregation level of all neighbourhoods that are accessible from each CBG in $\textbf{N}_s$, within $t$ minutes: 

\begin{equation}
    NA_{transit}(s,t) = \frac{\sum\limits_{n_s}^{\textbf{N}_s} \sum\limits_{n'}^{\textbf{N'}_{n_s}} ICE_{res}(n')}{\sum\limits_{n_s}^{\textbf{N}_s} |\textbf{N'}_{n_s}|}
    \label{eq:transit_seg}
\end{equation}

where $NA_{transit}$ is the neighbourhood accessibility when travelling by public transit. Equation \ref{eq:transit_seg} conveys the average socioeconomic profiles of neighbourhoods to which transit systems provide access, for CBGs belonging to a particular segregation group, $s$. This is achieved by determining the average segregation level of areas a segregation group can reach via transit, within a given time frame. We can calculate the same metric, but with respect to driving times, by using $T_{driving}$ to compute the set of reachable neighbourhoods, for a given segregation group and time threshold. We refer to this measure of segregation in driving access as $NA_{driving}$. We visualise both metrics in Figure \ref{fig:structural_transit}, for time thresholds from 5 to 60 minutes, at 5 minute intervals. For every matrix, the top row illustrates the changes in segregation characteristics of neighbourhoods that are accessible by the highly-segregated, low-income neighbourhoods (HS-Lo), for various time thresholds. Meanwhile, the bottom row captures average segregation levels, based on which neighbourhoods the highly-segregated, high-income neighbourhoods can reach. The top row of matrices illustrates how segregation of accessible neighbourhoods changes for transit time thresholds, measured using $NA_{transit}$. Meanwhile the bottom row of matrices captures driving accessibility ($NA_{driving}$), and serves as a baseline for comparison, as driving times calculated with OSRM are void of any transit schedule, route, or traffic constraints that are within the GTFS data used to build the transit-pedestrian networks. It is apparent that, when comparing transit to driving access for each city, segregation values for neighbourhoods accessible by car converge to reflect the city's overall socioeconomic composition much quicker than their public transit counterparts. 

To some extent, we would expect transit segregation to have different values across segregation groups, especially for smaller time thresholds, as a reflection of spatial auto-correlation in residential segregation. However, the differences in transit segregation levels persists beyond 60 minute journeys for Dallas, Cincinnati and New Orleans, revealing apparent structural inequalities in the transit systems of those cities. The driving access matrices in Figure \ref{fig:structural_transit} emphasise the disparities in transit service, using driving times to convey the possibility for transit services to provide less segregated accessibility.

\subsection{Transit Use Segregation}\label{ssec:Experiential_Transit_seg}
We wrap up our analysis of urban segregation by analysing how the areas to which individuals travel can impact the level of segregation experienced when using the transit system. We proceed, utilising the transit-pedestrian networks to calculate the shortest route, within the transit system, between a neighbourhood and an amenity. Specifically, we sample $\textbf{M}_{n,a}$ points from CBG $n$'s geographic boundaries, resembling the origin coordinates of visitors 
The destination points are defined by the longitude and latitude coordinates of amenity $a$. 
We note that by defining the economic composition of visitors by sampling neighbourhood income distributions, we assume uniform use of the transit system across socioeconomic groups. Thus, transit use segregation becomes an artefact of four main mechanisms: (1) the neighbourhood's income distribution, (2) the amenities its residents visit, (3) the frequency with which residents visit said amenities, and (4) how the subset of the transit system its residents use to visit their amenities intersects with the transit use of residents from other neighbourhoods. While this approach does not account for the fact that some demographics may rely more on transit \cite{giuliano2005low,hu2021left}, we hypothesise that incorporating disparities in transit reliance would exacerbate the levels of transit segregation we identify under this uniform-use assumption.

We can define the set of edges in the transit layer of the transit-pedestrian network that visitor $c$, in $\textbf{C}_{n,a}^x$ traverses when moving from their sampled origin coordinate to amenity $a$ as $\textbf{P}_{n,a,c}^{x}$. When an individual travels on a transit edge, her socioeconomic background contributes to the level of segregation experienced by all travelers using that transit segment. Thus, we can define the economic composition of a transit edge, $e$, for an iteration $x$ as:
\begin{equation}
     \textbf{C}_{e}^x = \{ c\ |\ n \in \textbf{N}, a \in \textbf{A}, c \in \textbf{C}_{n,a}^x, e \in \textbf{P}_{n,a,c}^{x}\} 
\end{equation}

To calculate segregation at the transit edge-level, we define $\textbf{C}_{e,i}^x$ to reflect the set of individuals from the low, middle or high income group that travel on an edge $e$, where $i \in \{lo, mid, hi\}$. In this manner we can define the level of segregation experienced on an edge, $e$, as an average of edge-level segregation across all stochastic iterations:

\begin{equation}
    ICE_{edge}^x(e) =  \frac{1}{100}\sum\limits_{x=1}^{100}\frac{|\textbf{C}_{e,hi}^x| - |\textbf{C}_{e,lo}^x|}{|\textbf{C}_{e}^x|}
\end{equation}

\begin{figure}
    \centering
    \includegraphics[width=0.8\textwidth]{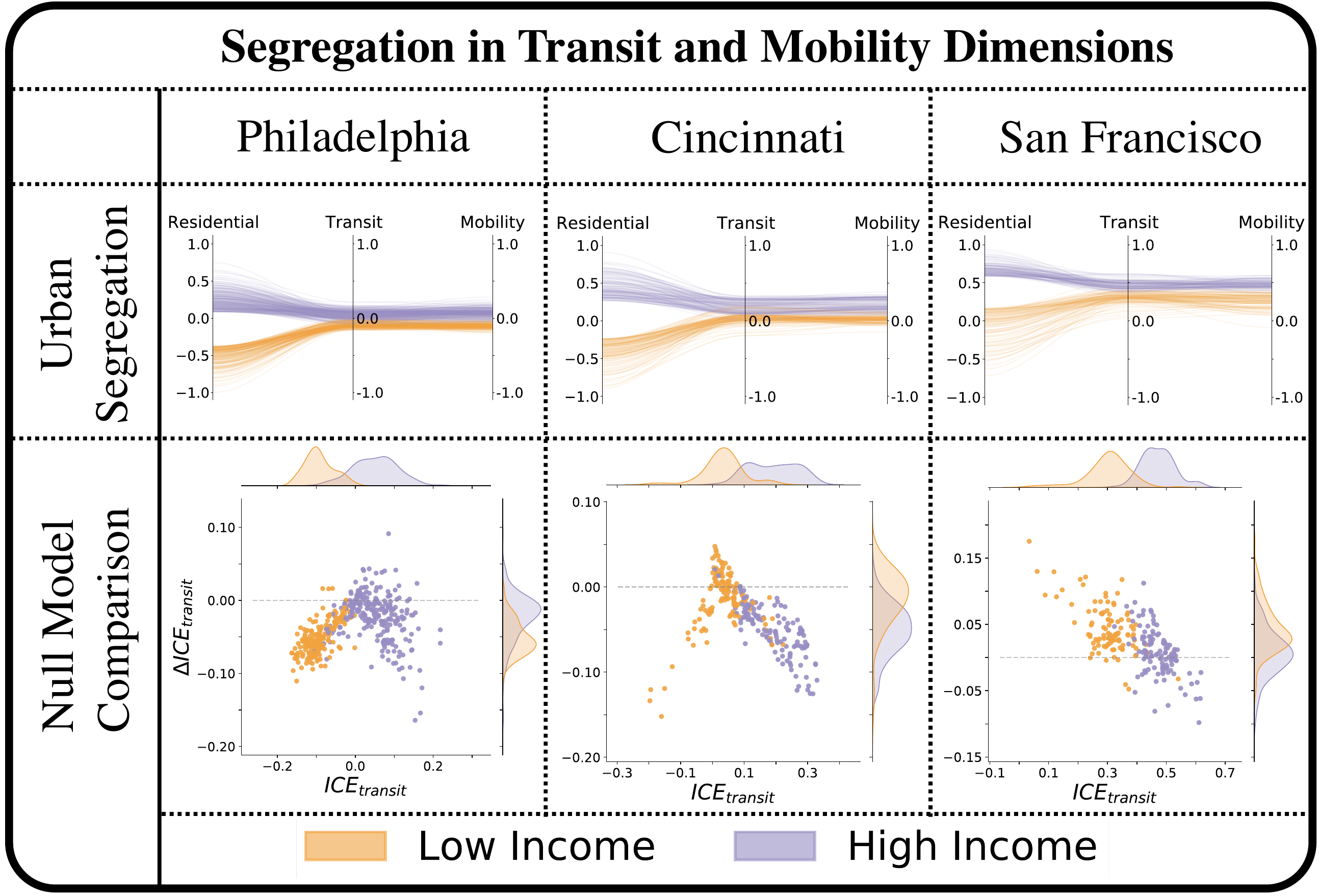}
    \caption{Urban segregation levels in Philadelphia, Cincinnati and San Francisco. The top row illustrates segregation across the residential, transit, and mobility domains, for the highly-segregated high and low-income groups, reflected by purple and orange, respectively. The bottom row shows how mobility segregation influences transit segregation by comparing empirical transit segregation to that of a null model, shown on the y-axis. The x-axis shows empirical levels of transit segregation as a baseline.}
    \label{fig:null_models}
\end{figure}

To compare how segregation levels change across the residential, amenity, and public transport domains, we aggregate edge-level transit segregation to the census block group level. For a given neighbourhood, $n$, we define the average segregation level experienced while using public transport, by residents in CBG $n$, as 
$\textbf{traveler transit segregation}$ (TTS): 

\begin{equation}
    TTS(n) = \overline{\textbf{T}_n} \textit{, where } \textbf{T}_n = \{ICE_{edge}(e)\ |\ a \in \textbf{A}, c \in \textbf{C}_{n,a}^x, e \in \textbf{P}_{n,a,c}^{x}\}
    \label{eq:ICE_transit}
\end{equation}
Thus, we have estimated how, residents in a given neighbourhood experience segregation from a residential (Eq. \ref{eq:ICE_res}), amenity (Eq. \ref{eq:TAS}), and public transport (Eq. \ref{eq:ICE_transit}) perspective. The top row in Figure \ref{fig:null_models} highlights how segregation levels persist across the three dimensions, focusing on the highly-segregated, low-income and highly-segregated, high-income segregation groups. The results for Dallas and New Orleans are included in Section \ref{si:transit_seg} the Supplementary Materials, however, we highlight the results for San Francisco, Cincinnati, and Philadelphia for brevity. Regardless of city-level economic composition, the parallel plots convey that segregation continues to exist in the transit and destination dimension, although to a lesser extent.  

To gain a deeper insight regarding the extent to which disparities in mobility destinations impact segregation while using the transit system, we develop a null model, which hypothesises that transit segregation is an artefact of disparities in the amenity landscape. To accomplish this, we retain the same distribution of trip counts across neighbourhoods in a city. However, we modify the destinations of every trip in the SafeGraph amenity visitations data, by randomly sampling a coordinate within a randomly sampled CBG. We construct a mobility matrix, $\text{M}_{n,a}^{\text{null}}$, using the sampled amenity destinations, and apply the same workflow to determine amenity and transit edge segregation. Again, to account for the stochasticity of sampling destinations, we perform this process for 100 iterations. Section \ref{si:null_seg} includes the parallel line plots for measured segregation in the null models that correspond to the top row of Figure \ref{fig:null_models}. 

 We can, then, compare segregation estimated from the empirical data, to that of the null model, which eliminates apparent disparities in the amenity dimension, illustrated in the bottom row of Figure \ref{fig:null_models}. 
 The x-axis and its corresponding distribution above it, convey the average level of segregation a neighbourhood's residents experience while using transit to satisfy their mobility demands. Thus, the x-axis more clearly visualises the Transit axis in the parallel line plots on the top row. Meanwhile, the y-axis, and respective distribution on the right, emphasises how transit segregation measures change when removing inequalities in amenity visitations and the amenity landscape. This is achieved by calculating $\Delta ICE_{transit}$, as defined in Equation \ref{eq:delta_ICE}, comparing a neighbourhood's empirical transit segregation to that measured in the null model. 
For most cities, we observe the majority of neighbourhoods having decreased levels of transit segregation when removing amenity visitation inequalities -- indicated by points falling below the dashed line. San Francisco remains an exception, implying that the amenity landscape and economic inequalities in amenity visitation shape the level of segregation individuals experience while using its transport system. Moreover, we observe the high income group experience larger decreases in transit segregation for most cities, as seen by the distribution on the y-axis. This suggests that the transit segregation experienced by individuals in highly-segregated, low-income neighbourhoods remains consistent, despite socioeconomic characteristics of their destinations. Additionally, the low magnitudes of $\Delta ICE_{transit}$ in the scatter plots elucidate how removing inequalities in amenity visitations and the amenity landscape does not significantly change segregation in the transit realm. We hypothesise that the identified transit segregation could, then, be a result of how the socioeconomic residential landscape intersects with the  transport service and layout.

Ultimately, we identify inequalities in how transit systems connect neighbourhoods from different socioeconomic backgrounds. We compare the average segregation of neighbourhoods that are reachable withing a given time, between trips taken using public transport versus cars. We note that San Francisco and Philadelphia allow residents from different segregation groups to reach a wider array of neighbourhoods within an hour long trip. Moreover, we stochastically model the transit lines individuals would use to satisfy their mobility demands. 
Finally, we test our empirical results against a null model to find that while disparities in amenity distribution and travel behaviour increase the level of economic concentration on transit lines, mobility and amenity inequalities do not fully account for the level of experienced transit segregation that we do identify.

\section{Discussion}\label{sec:discussion}

Inequalities in urban infrastructure can have a significant impact on perceived activity space and, consequently, travel behaviour. Urban analytics research aims to create more just cities by characterising these disparities, with recent attention focusing on accessibility measures \cite{smartcities4010006}. However, research in transport poverty highlights how lower income individuals tend to live in areas with higher levels of amenity accessibility \cite{allen2019sizing}. To understand how other forms of disadvantage can emerge, even in areas with high accessibility, we study urban inequalities regarding how transit provides access to neighbourhoods of different socioeconomic profiles, identifying how disparities in transit service can impact the experience of segregation throughout many urban dimensions.

This work puts forward a framework for defining inequalities in transit systems in terms of $\textit{where}$ transit provides access to and $\textit{how}$ individuals experience segregation while using transport. In doing so, our results reveal that residential segregation levels persist through other aspects of the urban experience, namely amenity visitations and transport usage. These results are consistent with research that shows residual effects of residential segregation in school, work, and mobility dimensions \cite{nieuwenhuis2021residential,delmelle2021poverty,silm2021relationship}. Furthermore, we show how highly segregated, low-income neighbourhoods tend to correct for their extreme levels of residential segregation through their mobility patterns. This is in line with findings that unveil demographic associations with social exploration of amenities \cite{moro2021mobility}. Bridgeport and San Francisco serve as two exceptions to this trend, where a subset of the neighbourhoods with low-income concentration tend to visit amenities with high levels of segregation. In these cases, it is imperative to develop adequate urban infrastructure, that is designed to benefit the disadvantaged groups that have low amenity accessibility \cite{allen2019sizing}. We also find that transit systems can hinder access to neighbourhood, limiting the potential of exposure to individuals from different backgrounds. These results underscore research that presents how mobility patterns in neighbourhoods with a high concentration of underprivileged demographics, be it immigrant or ethnic minorities, tend to be have more constrained activity spaces than their privileged counterparts \cite{hedman2021daily, silm2021relationship}. It is unclear whether mobility patterns are dictated by segregation in neighbourhood accessibility. What is apparent, however, is that by limiting exposure to different types of neighbourhoods, transit systems impose constraints on the activity space and urban experience of individuals, namely those without access to personal vehicles.

Limitations of this work include the assumptions that come with defining the economic composition of a neighbourhood's travellers regarding the neighbourhood's income distribution. Although it is striking that we identify inequalities under this assumption, which removes demographic mobility preferences within a neighbourhood, higher resolution mobility data can provide closer approximations of urban segregation. Thus, this work can be further developed to analyse how segregation experienced within transit lines is impacted by empirically informed levels of socioeconomic transit usage. Moreover, using higher-resolution mobility data, such as those that tag mobility trajectories with the associated demographics of the traveller, could shine light on further disparities in how transport and amenity landscapes intersect. Additionally, the proposed methodology can be applied to data spanning a larger time frame, to analyse temporal features of mobility and transit segregation. We emphasise that this framework can be applied to any region, given transit feeds for building transport networks and mobility data that includes or can be merged with demographic characteristics. By applying this workflow to  manner, we can have a clearer insight on how cultural differences in mobility patterns and the level of transit infrastructure can impact inequalities that residents experience in various facets of the urban environment.

 In essence, we consider segregation from multiple urban dimensions to highlight the benefit of analysing segregation as a spatio-temporal experience rather than a static variable. Moreover, identifying inequalities within transit systems is the first step in providing improved transit service, particularly to individuals from especially vulnerable demographics. By studying segregation from multiple perspectives, we can observe whether mobility is used as a tool to try and overcome residential segregation.

\putbib

\end{bibunit}
\newpage

\begin{bibunit}
\titleformat{\part}[display]
{\normalfont\LARGE\bfseries\centering}{}{12pt}{Supplementary 
Materials}
\titleformat{\subparagraph}[runin]
{\normalfont}
{\thesection}{.5em}{}
\titlespacing{\section}{10pt}{10pt}{4pt}
\titlespacing{\subsection}{10pt}{10pt}{4pt}

\part{}

\renewcommand {\thesection}{S\arabic{section}}
\renewcommand{\thetable}{S\arabic{table}}
\renewcommand {\thefigure}{S\arabic{figure}}

\section{Details on the Index of Concentration at the Extremes \label{si:ice_details}}
\subsection{Correlation of ICE with Different Segregation Metrics\label{si:sec_ice_seg_corr}}

Here we compare how socioeconomic segregation, measured using the Index of Concentration at the Extremes, compares to other metrics of segregation. To avoid repetition, we group cities by their mean ICE value, where each group is a row of scatter plots. Thus, the first row depicts ICE correlations for cities with the lowest mean segregation, while the last row does so for cities with the highest mean segregation. The blue scatter plots capture segregation correlations at the census tract level. Census block groups (CBGs) are the highest resolution for which household income distributions can be openly accessed. Conventional segregation measures estimate inequality in an area using the distribution of individuals across its subareas. Thus, conventional measures such as dissimilarity (capturing the unevenness dimension) and isolation (depicting the exposure dimension) are constrained to the census tract level, which is one step larger than CBGs in terms of geographic boundaries.
We see that Dissimilarity and ICE don't exhibit any apparent relationship, while isolation expressed curves similar to that of the logarithmic function. Thus isolation is more useful at capturing disparities in highly segregated low-income tracts. However, isolation vales are high for positive ICE values, making it hard to identify segregation in terms of high-income concentration. The Mutual Information Index, derived from information theory, most closely captures the inequalities that ICE does, except it does so from a range of 0 to 1 \cite{theil1971note,reardon2002measures}. Social distance introduced by \cite{xu2019quantifying}, uses a fractional rank-based approach to measure residential inequalities. We see a consistently parabolic relationship between this measure and ICE, which highlights the utility of ICE in distinguishing between segregation of the most and least privileged demographics. However, future work could focus on exploring how the use of different segregation metrics impacts identified segregation levels throughout urban dimensions.

\begin{figure}[H]
  \centering
  \includegraphics[width= \textwidth]{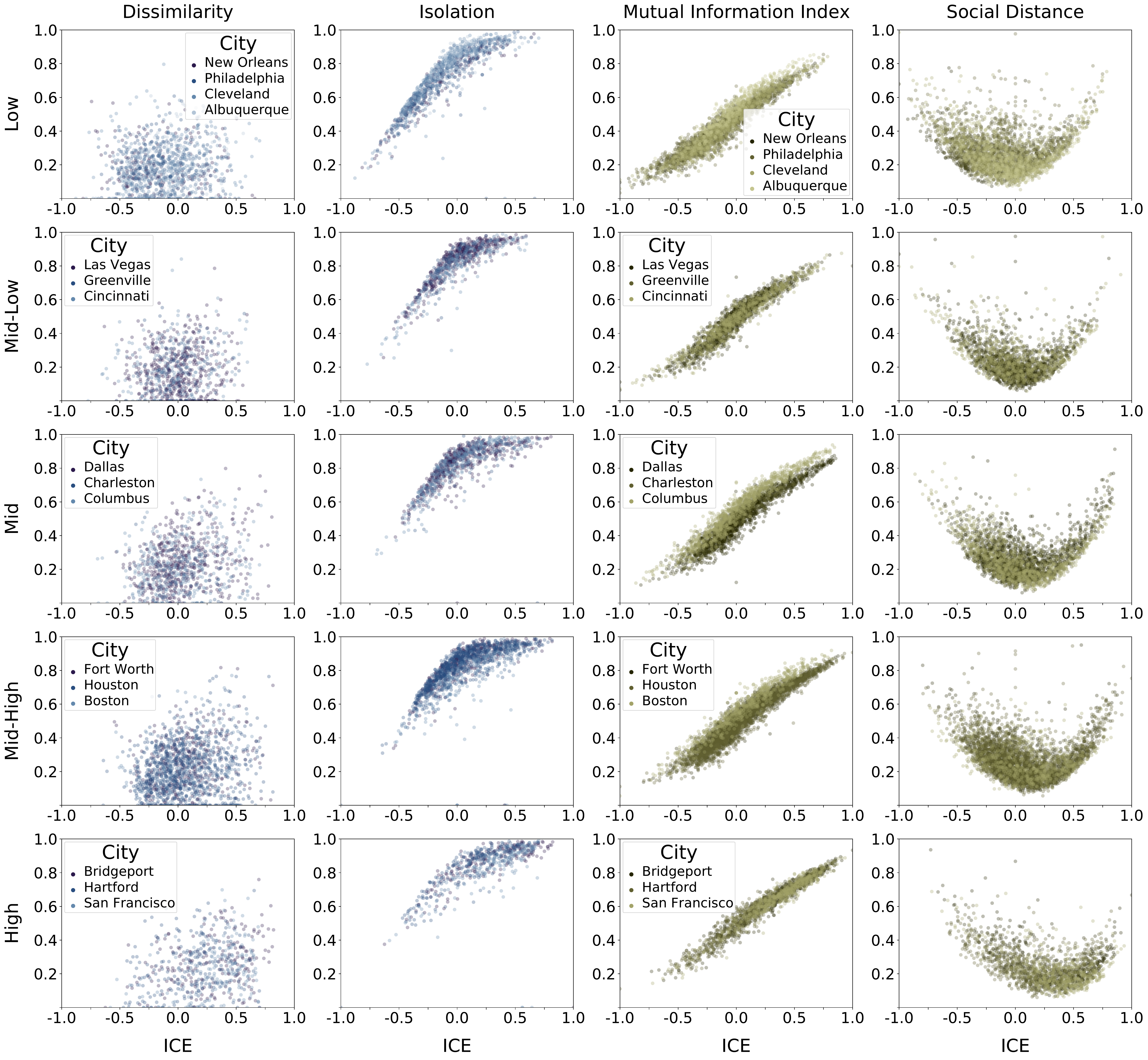}
  \caption{Correlations between residential socioeconomic segregation (ICE) and other segregation metrics. ICE is compared to the index of Dissimilarity, Isolation, Mutual Information and Social Distance in the first, second, third, and fourth column, respectively. The x-axis of each scatter plot reflects ICE values, while the y-axis depicts values for the respective segregation metric to which ICE is being compared. Blue plots reflect comparisons at the census tract level, while green plots captures segregation for census block groups. Each row reflects the ICE correlations for a group of cities that are paritioned based on their mean ICE values.}
  \label{si:ice_seg_corr}
  
\end{figure}

\subsection{The Impact of Defining Low and High Income Cutoffs\label{si:ice_seg_cutoffs}}
Here we explore the impact of using different income percentiles to define the number of low income and high income households in a neighbourhood. Conventionally, when defining ICE, low and high income groups are defined by the 20th and 80th percentile of the income distribution, respectively. In the context of the Median Household Income data from the US Census Bureau American Community Survey, income distributions are split into 16 categories. The 20th and 80th percentile correspond to the 3 lowest and 3 highest brackets, representing households earning less than \$20,000 and those earning more than \$125,000. Figure \ref{si:ice_seg_cutoffs} shows how the distribution of ICE values in a city changes when we shift the boundaries that define low and high income groups. We see that by creating broader definitions of what it means to be an extremely low or high-income household, neighbourhoods tend to have a more distributed range of ICE values. Moreover, broader definitions lead to an increase in cities' median ICE values.
\begin{figure}[H]
  \centering
  \includegraphics[width=1 \textwidth]{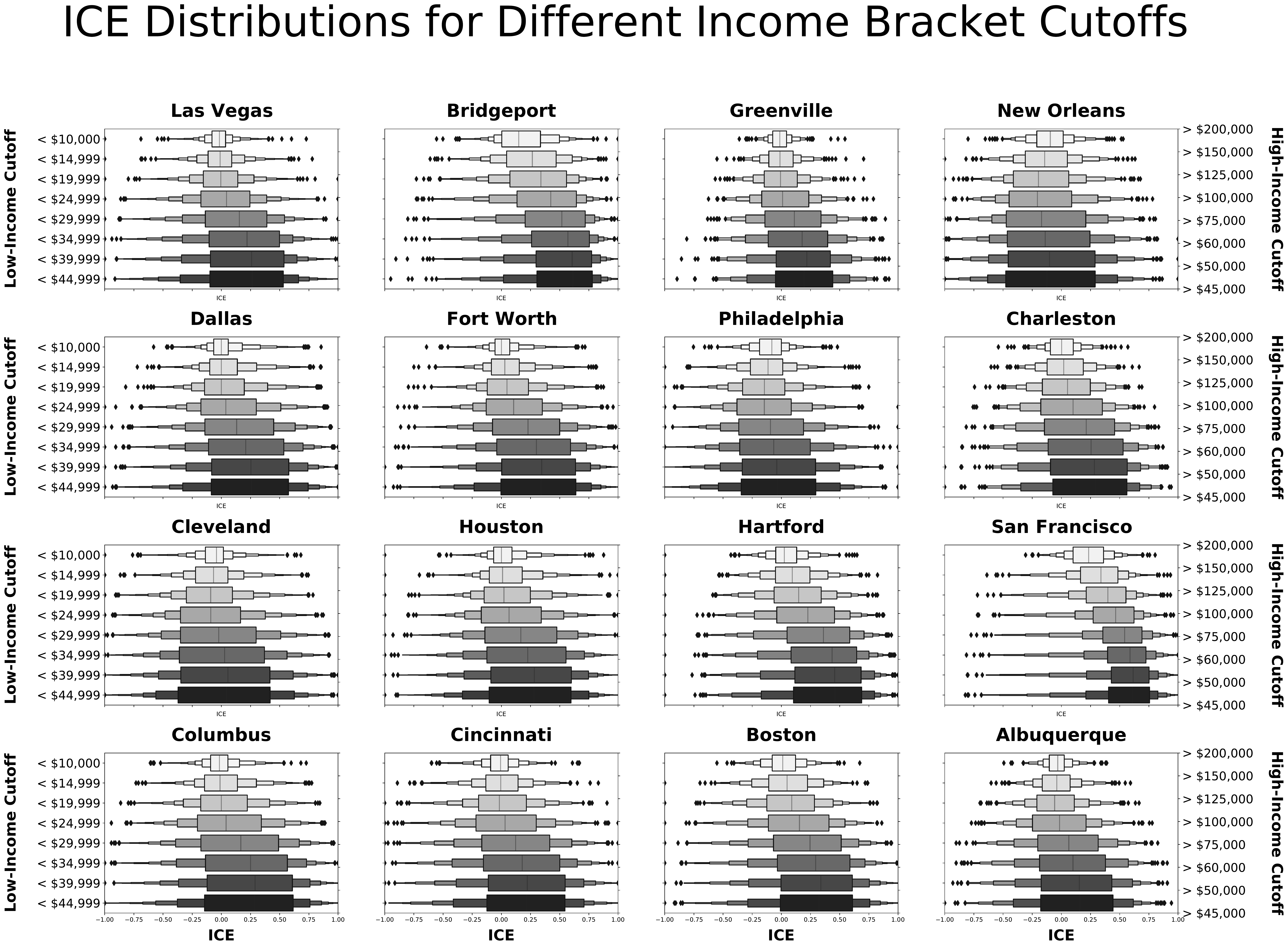}
  \caption{Each scatterplot reflects distribution of ICE values in a city at the census block group level. Each box-plot refers to the ICE distribution (x-axis) for a particular definition of what constitutes a low and high income household. The definition of low-income cutoffs is on the left axis of the figure, while that of the high-income is on the right axis.}
  \label{fig:ice_seg_corr}
  
\end{figure}

\subsection{Defining Segregation Groups \label{si:seg_groups}}

\begin{figure}[H]
  \centering
  \includegraphics[width=1 \textwidth]{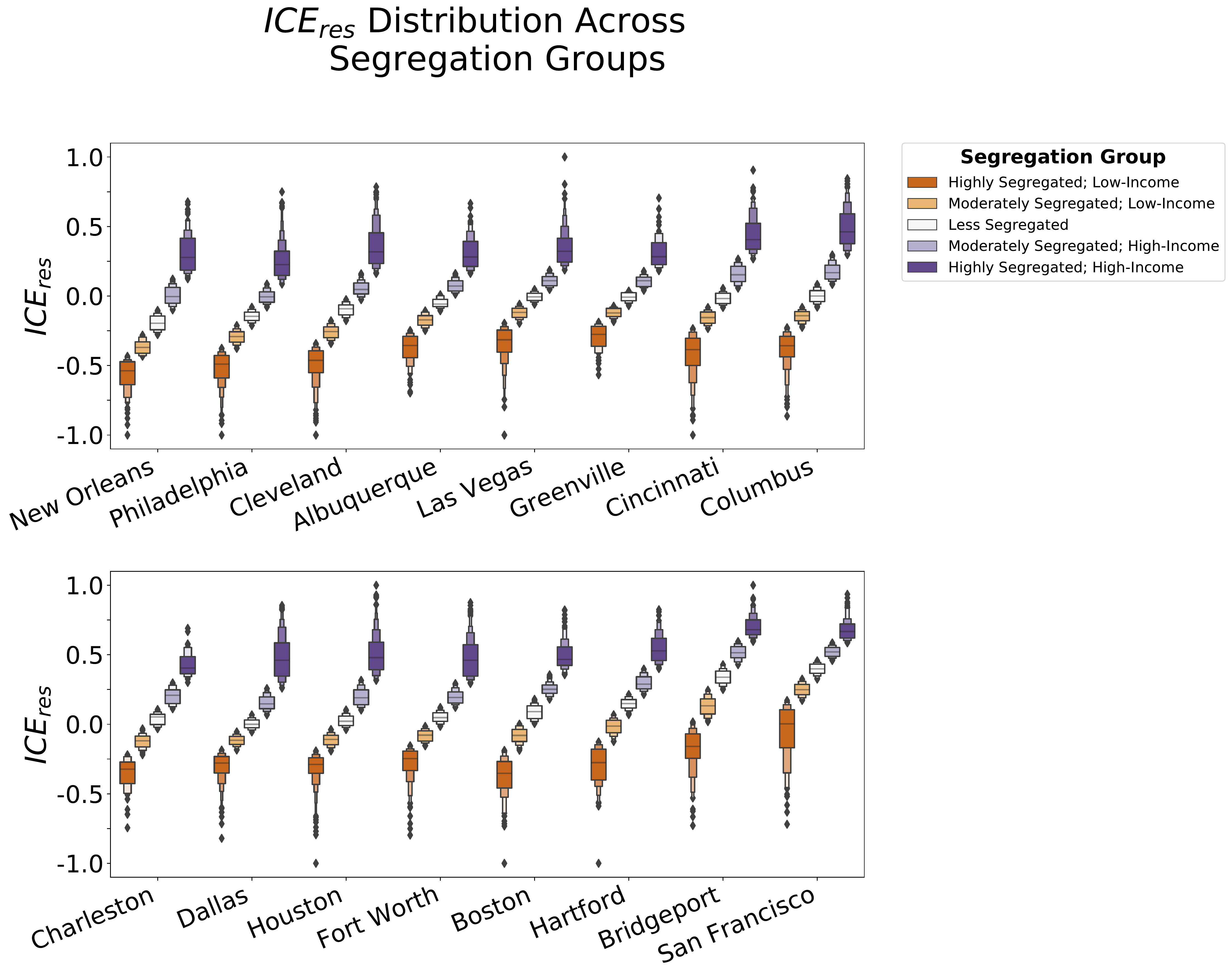}
  \caption{ICE distribution values within each segregation group, for a given city. Orange box plots depict low-income concentration segregation groups, while purple reflects segregation groups of high-income concentration. Darker colors capture higher segregation. The y-axis shows the ICE values of census block groups belonging to a particular segregation group.}
  \label{si:ICE_seg_grp}
  
\end{figure}

In the main manuscript, we define 5 segregation groups: (1) highly-segregated; low-income (\textbf{HS-Lo}) (2) moderately-segregated; low-income (\textbf{MS-Lo}) (3) less-segregated (\textbf{LS}) (4) mildly-segregated; high-income (\textbf{MS-Hi}) and (5) highly-segregated; high-income (\textbf{HS-Hi}). We create these groups by splitting neighbourhoods intwo 5, equally-sized categories, using levels of residential segregation for partitioning. Figure \ref{si:ICE_seg_grp} illustrates, for each city, the distribution of ICE values that belong to each group. Through this figure, it becomes clear that a HS-Lo segregation group in one city does carry the same meaning when applied to another city, due to differences in the economic composition of cities.

\section{Details on Transport Data}
\subsection{Creating Transit-Pedestrian Networks\label{si:transit_nx}}

We construct transit-pedestrian networks for a given city using General Transit Feed Specification data and extracts from OpenStreetMap. Nodes represent transit stops and street nodes. Thus, there are four types of edges: (1) edges connecting two transit stops (2) edges connecting two street nodes (3) edges connecting a street node to a transit node and (4) edges connecting at transit node to a street node. Table \ref{si:transit_nx_table} includes details regarding the transit layer of the transit-pedestrian network as well as GTFS data specifications, such as the number of unique routes and unique trips in the GTFS data, given a specific time frame. We construct our networks assuming transit use on Monday from 08:00 to 10:30. The directed transit-transit edges and edge weights are created from the GTFS operational schedule, with respect to the user-specified day and time window. The directed transit-pedestrian edges connect every transit stop to its closest pedestrian node, deriving travel time using distance and an assumed walking speed of three miles per hour. Meanwhile, the undirected pedestrian-pedestrian edges use the distance between each node pair in the OSM network to weight the edges with travel time, again assuming a walking speed of three miles per hour. Finally, the directed pedestrian to transit edges, similar to the transit to pedestrian edges, maps every transit node to its nearest pedestrian node. However the pedestrian to transit edges account for wait time at the transit stops based on average transit headway -- the amount of time between transport vehicles at a stop. Thus, the pedestrian to transit edges differ from the transit to pedestrian edges, in that they account for both the walking time from pedestrian to transit nodes and the wait time at the transit stop. Further details regarding the construction of the transit-pedestrian network can be found in \cite{blanchard2017urbanaccess}.

\begin{table}[]
    \centering
    
    \begin{tabular}{lrrrr}
    \toprule
              City &  \# Nodes &  \# Edges &  \# Trips &  \# Routes \\
    \midrule 
        Greenville &      462 &      917 &       35 &        12 \\
        Charleston &     1067 &     5842 &      278 &        23 \\
       Albuquerque &     2114 &     7858 &      224 &        25 \\
          Columbus &     3441 &    13222 &      292 &        25 \\
        Fort\ Worth &     1946 &     8195 &      354 &        36 \\
        Bridgeport &     2433 &     8380 &      311 &        37 \\
       New\ Orleans &     2926 &    10679 &      328 &        37 \\
       Gainesville &     1913 &    15419 &      797 &        42 \\
         Cleveland &     6630 &    53259 &     1129 &        45 \\
        Cincinnati &     4692 &    19481 &      440 &        47 \\
         Las\ Vegas &     7987 &    35812 &     1111 &        76 \\
          Hartford &     6493 &    20464 &      682 &        93 \\
            Dallas &     7773 &    37068 &     1150 &       102 \\
           Houston &    11632 &    58638 &     1317 &       112 \\
      Philadelphia &    12062 &    94170 &     2430 &       125 \\
     Dan\ Francisco &     8682 &    80114 &     3115 &       156 \\
            Boston &     6798 &    46883 &     3067 &       204 \\
    \bottomrule
    \end{tabular}
    \caption{Transit network properties for 16 US cities}
    \label{si:transit_nx_table}
\end{table}

\section{Correlations between Residential and Mobility Segregation}

Table \ref{si:res_mob_corr} conveys the significant Pearson correlations between residential and mobility segregation for each of 16 cities that we analyse.

\begin{table}[]
    \centering
    
    \begin{tabular}{lcc}
        \toprule
        \multicolumn{1}{c}{City}  & \multicolumn{2}{c}{$ICE_{res}\ \alpha\ ICE_{TAS}$} \\
        \cmidrule{2-3}
        {} &    {$r^1$} &       $p$ \\
        \midrule
        Las Vegas     & \textbf{7.526e-01***} & 1.128e-234 \\
        Charleston    & \textbf{7.640e-01***} &  7.596e-46 \\
        Greenville    & \textbf{7.788e-01***} &  3.655e-53 \\
        Fort Worth    & \textbf{7.969e-01***} & 8.735e-259 \\
        Cincinnati    & \textbf{7.976e-01***} & 2.872e-154 \\
        Columbus      & \textbf{8.001e-01***} & 1.068e-237 \\
        Houston       & \textbf{8.051e-01***} &  0.000e+00 \\
        Albuquerque   & \textbf{8.116e-01***} & 1.164e-102 \\
        Dallas        & \textbf{8.197e-01***} &  0.000e+00 \\
        Philadelphia  & \textbf{8.249e-01***} &  0.000e+00 \\
        Boston        & \textbf{8.262e-01***} & 3.681e-159 \\
        Hartford      & \textbf{8.378e-01***} & 3.829e-176 \\
        New Orleans   & \textbf{8.547e-01***} & 8.902e-141 \\
        Cleveland     & \textbf{8.597e-01***} &  0.000e+00 \\
        Bridgeport    & \textbf{8.661e-01***} & 6.295e-198 \\
        San Francisco & \textbf{9.107e-01***} & 2.736e-223 \\
        \bottomrule
        \multicolumn{3}{l}{$^1$ *$p<0.05$; **$p<0.01$; ***$p<0.001$} \\
        
    \end{tabular}
    \caption{Pearson correlation coefficients between socioeconomic residential segregation of a neighbourhood and the experienced segregation of its residents based on the amenities they are visiting, for 16 US cities.}
    \label{si:res_mob_corr}
\end{table}


\section{Transport Segregation}
\subsection{Transit Segregation for Dallas and New Orleans \label{si:transit_seg}}
\begin{figure}[H]
  \centering
  \includegraphics[width=0.7 \textwidth]{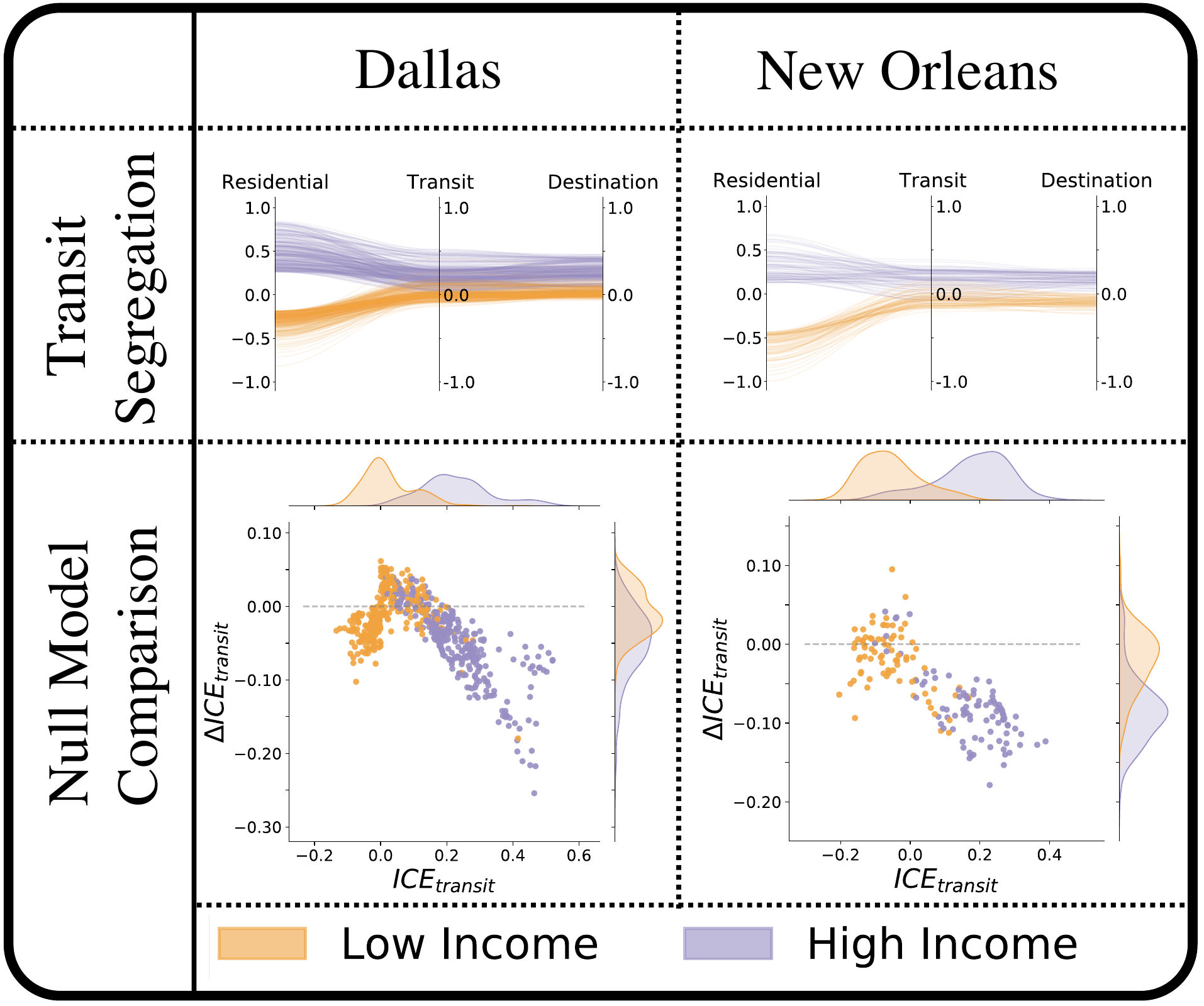}
  \caption{Segregation levels in Dallas an New Orleans for different urban contexts. The top row visualises segregation across the residential, transit, and mobility domains, for the HS-Hi and HS-Lo groups, reflected by purple and orange, respectively. The bottom row reflects how segregation in mobility impacts transit segregation by comparing the empirical data to that of a null model ($\Delta ICE_{transit}$). The y-axis illustrates these changes, while the x-axis depicts empirical $ICE_{transit}$ to provide context for the $\Delta ICE_{transit}$ values.}
  \label{si:dallas_neworleans}
  
\end{figure}

We show transit segregation and its relation to mobility segregation in Dallas and New Orleans in Figure \ref{si:dallas_neworleans}. This was excluded from the main manuscript due to space constraints, but follows similar trends of persistent segregation in multiple urban domains. Moreover, as discussed in the paper, transit segregation in both cities is impacted by mobility segregation (as seen by the quantity of neighbourhoods with negative $\Delta ICE_{transit}$ values). Lastly, we note the smaller $\Delta ICE_{transit}$ values for the highly-segregated, high-income (HS-Hi) segregation group in the scatter plots. These values imply that the segregation levels that the HS-Hi group experience on transit routes is more influenced by mobility segregation than the HS-Lo group. 

\subsection{Null Model Segregation Results \label{si:null_seg}}
\begin{figure}[H]
  \centering
  \includegraphics[width=1 \textwidth]{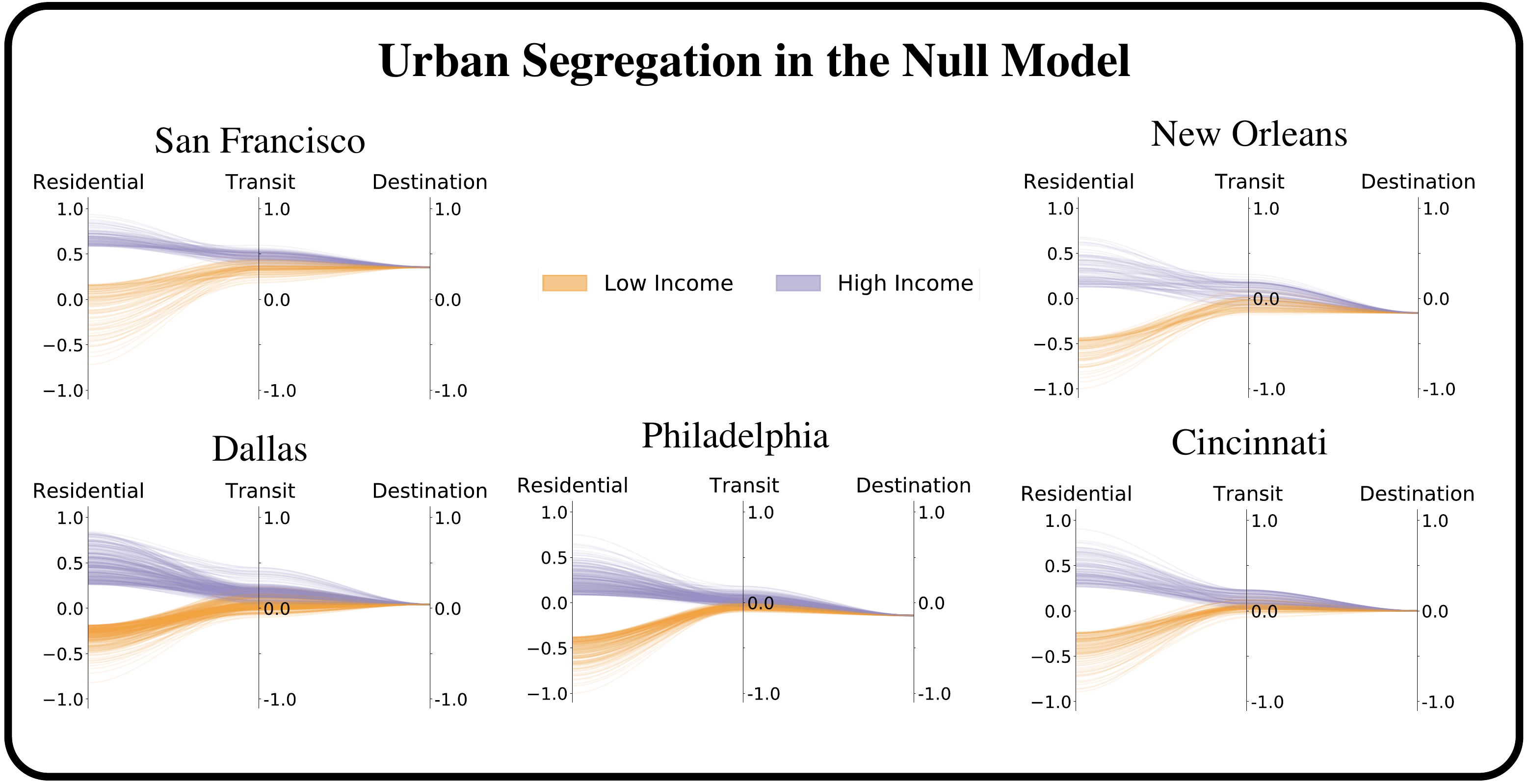}
  \caption{Changes in segregation between the residential, transit, and mobility domains, using mobility generated for the null model comparison. Shown for 5 US Cities, for which the null model simulates mobility patterns to uniform destination. Where orange is the HS-Lo segregation group and purple represents the HS-Hi segregation group.}
  \label{si:transit_seg_null_model}
  
\end{figure}

This section serves to complement Figure \ref{fig:null_models} in this main manuscript. Figure \ref{si:transit_seg_null_model} shows how measures of segregation change across urban dimensions, using amenity visitations that are generated for the null model rather than empirical mobility patterns. We observe that while disparities in amenity segregation are removed, as seen by the converging lines in the Destination axis, segregation in the transit dimension continues to exist. Moreover, mobility segregation corresponds to values that are reflective of the city's overall socioeconomic composition.

\putbib

\end{bibunit}

\end{document}